\documentclass{article}

\usepackage{arxiv}

\usepackage[utf8]{inputenc} 
\usepackage[T1]{fontenc}    
\usepackage{hyperref}       
\usepackage{url}            
\usepackage{booktabs}       
\usepackage{amsfonts}       
\usepackage{nicefrac}       
\usepackage{microtype}      
\usepackage{lipsum}
\usepackage{amsmath}
\usepackage{graphicx}
\usepackage[section]{placeins}
\usepackage[numbers,sort&compress]{natbib}  

\title{Reducing the Dimensionality of Optimal Experiment Design for Magnetic Resonance Fingerprinting}

\author{
  Nikolai J. Mickevicius, Ph.D.\\
  Department of Radiation Oncology\\
  Medical College of Wisconsin\\
  Milwaukee, WI 53226 \\
  nmickevicius@mcw.edu\\
  \And
  Andrew S. Nencka, Ph.D.\\
  Department of Radiology\\
  Medical College of Wisconsin\\
  Milwaukee, WI 53226 \\
  \And
  Eric S. Paulson, Ph.D.\\
  Department of Radiation Oncology\\
  Medical College of Wisconsin\\
  Milwaukee, WI 53226 \\
}

\begin{document}
\maketitle

\begin{abstract}
Nuclear magnetic resonance signal dynamics as described by the Bloch equations are highly complex and often are without closed form solutions. This is especially the case for quantitative magnetic resonance fingerprinting (MRF) scans in which acquisition parameters are varied to efficiently probe the parameter space. As previously demonstrated, optimization of the pattern in which the MRI acquisition parameters are varied relative to the variance in the target quantitative parameters can improve experiment design. This process, however, relies on large scale non-linear optimizations with hundreds to thousands of unknowns. As such, the numerical optimization is extremely time consuming, highly sensitive to guesses, and prone to getting caught in local minima. Here, we describe a method to reduce the complexity of the optimal MRF experiment design by constraining the solutions to span a predetermined low-dimensional subspace. Compared with standard flip angle patterns in calibration phantom experiments, precision in $T_2$ can be increased by optimizing as few as 8 coefficients in less than one minute of computation time.
\end{abstract}

\keywords{Relaxometry \and MR Fingerprinting \and Optimization \and Cramér-Rao Lower Bound \and Extended Phase Graphs}

\section{Introduction}
The development of fully quantitative imaging protocols has been the goal of many research groups since the inception of MRI.  Exam durations, limited by patient comfort and clinical throughput, are not typically long enough to accommodate most desired quantitative MR imaging scans. A large number of methods have been developed over the years to rapidly and accurately compute $T_1$ and $T_2$ relaxation times\cite{Poon,Deoni2003,Schmitt2004,Warntjes2007,Doneva2010}. In theory, with these measurements, synthetic images from most common pulse sequences may be calculated for radiological diagnoses (e.g. $T_1$-weighted, $T_2$-weighted, FLAIR, STIR, etc.) and the quantitative information necessary for robust MRI biomarker research is obtained\cite{Tanenbaum2017}.

Recently, magnetic resonance fingerprinting (MRF) has largely impacted the quantitative MRI field\cite{Ma2013}. In this type of acquisition, the flip angle, echo time (TE), repetition time (TR), and k-space coverage can be varied pseudo-randomly throughout the course of a rapid gradient echo scan. The basic idea is that the dynamics of the magnetization in different tissues will evolve with the changes in acquisition parameters in a unique way. A time course of undersampled images, each frame typically reconstructed from one interleaf of a variable density spiral (VDS), is utilized to measure a signal evolution within each voxel. The measured signal evolutions are then compared to a dictionary which contains the ideal signal evolutions calculated from Bloch equation simulations\cite{Bloch1946,Jaynes1955} or via extended phase graphs (EPG)\cite{Weigel2015}, for many combinations of tissue parameters including, but not limited to $T_1$, $T_2$, proton density, off-resonance, diffusion coefficient, and magnetization transfer ratio. The tissue parameters which simulated the time course best-matched to the measured data are assigned to the appropriate imaging voxel. 

Rather than varying acquisition parameters in a pseudo-random or arbitrarily chosen fashion\cite{Ma2013,Jiang2015}, the optimization of flip angle and repetition time patterns for MRF has become an active area of research\cite{Zhao2019,Asslander2019}. Such optimizations reduce the variance in the unbiased estimation of $T_1$ and $T_2$. Since MRF acquisitions can consist of several thousand of RF excitations, the scale of a optimization algorithm to determine the best acquisition parameters is immense. Such numerical optimizations typically change flip angle and/or repetition time patterns until a statistical measure of the variance in the estimation of the quantitative parameters is minimized. For relatively short MRF acquisitions (i.e., $N_{RF}=400$), published optimizations can take on the order of 290 minutes\cite{Zhao2019}, and the end result is highly dependent on the initial guess for the flip angle pattern. Thus, if multi-start optimization methodologies are employed in pursuit of a global minimum, the optimization could take days to complete without any guarantee of convergence upon a true global minimum. 

If the dimensionality of MRF optimization could be reduced, it would open the door to a more efficient search for the best possible acquisition parameters for the estimation of quantitative MRI parameters from highly efficient acquisitions. Further, the development of a reduced dimensionality optimization framework for MRF could also allow for scanner-specific adjustments to the acquisition parameters to compensate for system imperfections such as $\Delta B_0$, $\Delta B_1^+$, or eddy current build up, for example. Thus, in this work, we aim to develop such a framework for gradient spoiled (i.e., FISP) MR fingerprinting acquisitions\cite{Jiang2015}. The goal is to test the hypothesis that similar reductions of the variance in the unbiased estimation of $T_1$ and $T_2$ with the reduced dimensionality approach compared with full scale optimizations. We employ small number of partially overlapping Gaussian functions as a basis to represent a smoothly varying flip angle pattern. Numerical optimizations of the basis function coefficients to minimize the variance in the estimation of $T_1$ and $T_2$ are performed. An assessment of the performance of these optimized patterns was performed quantitatively in calibration phantom experiments.

\section{Theory and Methods}
\label{sec:headings}

\subsection{Cramér-Rao Bound}
Let us consider a tissue with relevant tissue properties of longitudinal relaxation ($T_1$), transverse relaxation ($T_2$), and proton density ($M_0$). For a given tissue, $\boldsymbol{\theta}_i=\left[T_1, T_2, M_0 \right]^T$, the variance in the estimation of the unknown tissue properties are contained in the inverse of the Fisher information matrix, $\mathbf{F}\in \mathbb{C}^{3 \times 3}$. The Fisher information matrix is defined as $\mathbf{F}_{i,j}=\mathbf{b}_i^H\mathbf{b}_j/\sigma ^2$ where $\mathbf{b}_i$ represents the partial derivative of the time dependent signal evolution with respect to relevant tissue properties:

\begin{equation}\label{eq:b_derivs}
    \mathbf{b}_1=\frac{\partial \widetilde{\mathbf{F}}_0}{\partial T_1}
    \qquad
    \mathbf{b}_2=\frac{\partial \widetilde{\mathbf{F}}_0}{\partial T_2}
    \qquad
    \mathbf{b}_3=\frac{\partial \widetilde{\mathbf{F}}_0}{\partial M_0}
\end{equation}

The vector of measurable signal at all echo indices is given by, $\widetilde{\mathbf{F}}_0$, which uses the notation from the extended phase graph formalism described in the Appendix. The Cramér-Rao bound (CRB) describes a lower bound on the variance in the unbiased estimation of all quantitative parameters of interest\cite{Rao1992}. The relative CRB (rCRB) given input $\sigma^2$ is defined as follows, where $w_{T_1}$, $w_{T_2}$, and $w_{M_0}$ are used to apply different weights to each of the relevant quantitative parameters. 
\begin{equation}\label{eq:crb}
    rCRB=
    w_{T_1}M^2_0\frac{\left(\mathbf{F}^{-1}\right)_{1,1}}{\sigma^2 T^2_1} + 
    w_{T_2}M^2_0\frac{\left(\mathbf{F}^{-1}\right)_{2,2}}{\sigma^2 T^2_2} + 
    w_{M_0}\frac{\left(\mathbf{F}^{-1}\right)_{3,3}}{\sigma^2}
\end{equation}

\subsection{Optimal Experiment Design for MRF}

The primary goal of MRF experiments is to quantitatively map both $T_1$ and $T_2$ with high levels of precision. The goal of optimal MRF experiment design, in general, is to choose a pattern of acquisition parameters to minimize Equation \ref{eq:crb} using numerical methods. In this work, we will focus on the optimization of flip angle schedules for FISP MRF to minimize Equation \ref{eq:crb}. We will keep the repetition time (TR) constant because 1) in our experience, optimizing the TR simultaneously with flip angle does not provide enough improvement in rCRB to warrant doubling the number of unknowns and 2) subjects scanned with variable TR MRF acquisitions frequently express discomfort over the unique acoustic noise. Thus, for an MRF experiment consisting of $N_{RF}$ consecutive FISP excitation, readout, and spoiling repetitions requires optimization of a vector of flip angles $\boldsymbol{\alpha}\in \mathbb{R}^{N_{RF}}$. As shown in Zhao et. al., non-linear optimization algorithms such as sequential quadratic programming (SQP) can be used to solve for $\boldsymbol{\alpha}$ to minimize Equation \ref{eq:crb}\cite{Zhao2019}. Also described in \cite{Zhao2019} is a heuristic that smoothly varying flip angle patterns produce more favorable MRF results, likely because they allow a separability of noise from k-space undersampling and noise-like changes in signal intensity due to magnetization dynamics. Here, we use this heuristic as motivation to represent favorable, smoothly varying flip angle schedules using a low-dimensional set of basis functions. We hypothesize that this will allow for a reduction in the number of variables to optimize by factors while still providing substantial reductions in rCRB compared to published flip angle patterns.

\subsection{Numerical Optimization of Flip Angle Patterns}

Although Fourier or spline functions could be used to represent a smoothly varying function, we chose the use of a linear combination of partially overlapping, equally-spaced Gaussian functions as the basis upon which flip angle patterns for a FISP MRF acquisition will be calculated. A basis of $K$ Gaussian functions can be stacked into a matrix $\mathbf{A} \in \mathbb{R}^{N_{RF} \times K}$. The $p$th column of $\mathbf{A}$ is given by $\mathbf{a}_p = e^{(\mathbf{n}-\mu_p)^2/2s^2}$. Here, $\mathbf{n} \in [1, 2, \dots, N_{RF}]^T$, $\mu_p$ is the desired center of the Gaussian peak, and $s = \frac{1}{2}(\mu_2 - \mu_1)$. The optimization of the flip angle pattern then reduces to the non-linear optimization of a vector with $K$ coefficients, $\mathbf{x} \in \mathbb{R}^K$, such that $\mathbf{A}\mathbf{x}$ minimizes Equation \ref{eq:crb}. A constraint that $0^\circ \leq \mathbf{A} \mathbf{x} \leq 70^\circ$ was used to maintain linearity of the asymmetric sinc RF pulse used for excitation. Details of the acquisition include $N_{RF}=800$, a constant TR of 8 ms, TE of 2.4 ms, and inversion time of 20 ms. 

\begin{figure}[h]
\centering
\includegraphics[width=0.6\textwidth]{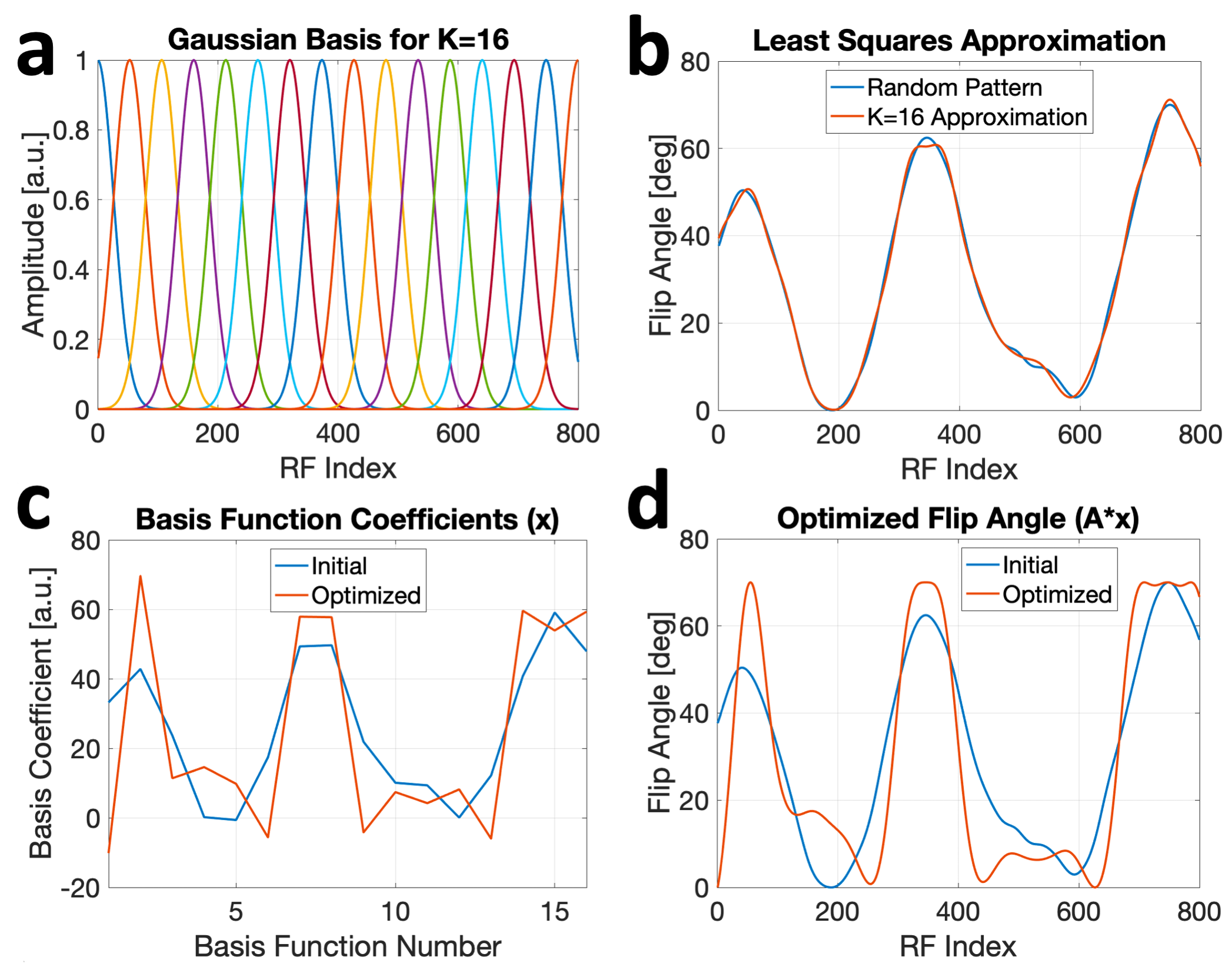}
\caption{(a) Example set of Gaussian basis functions for $K=16$. (b) Least squares approximation of a random initial flip angle pattern using Gaussian basis. (c) Initial and numerically optimized basis function coefficients. (d) The initial flip angle pattern and optimized flip angle pattern obtained by projecting optimized coefficients through the basis.}
\label{fig:method}
\end{figure}

Patterns were calculated for values of $K$ between 4 and 28 in steps of 4. For each $K$, 10 optimizations were performed with different initial conditions. Uniform noise was filtered with random moving average filters and scaled to span a range between $0^\circ$ and $70^\circ$ to obtain a random flip angle pattern, $\mathbf{y}$. The initial guess of $\mathbf{x}$ for each $K$ and initial condition was obtained via $\mathbf{x}_0 = (\mathbf{A}^T \mathbf{A})^{-1}\mathbf{A}^T \mathbf{y}$. A constrained SQP algorithm from MATLAB (The MathWorks, Natick, MA) was used to generate optimal flip angle patterns. To reduce variance in the estimation of $T_1$ and $T_2$, $w_{T_1}$ was set to 1, $w_{T_2}$ was set to 1, and $w_{M_0}$ was set to zero. The code to calculate the vectors in Equation \ref{eq:b_derivs}, as detailed in the Appendix, was written in C with a MEX interface to MATLAB. The best performing flip angle schedule for each value of $K$ was then selected for quantitative \textit{in silico} and phantom MRF experiments. A maximum of 700 iterations of the SQP algorithm were used with a tolerance of $1\times 10^{-6}$. The sum of Equation \ref{eq:crb} for two tissues, $\boldsymbol{\theta}_1=[785, 65, 1]$ and $\boldsymbol{\theta}_2=[1200, 110, 1]$, was minimized during optimization. The $T_1$ and $T_2$ times here are in milliseconds. A summary of the proposed optimization methodology can be seen in Figure \ref{fig:method}.

\subsection{MRF Dictionary Calculation}
MRF signal evolution dictionaries were calculated for the best performing flip angle patterns for each value of $K$ and published flip angle patterns. The same extended phase graph (EPG) code base used to calculate the vectors in Equation \ref{eq:b_derivs} was used to simulate tissue fingerprints for a range of $T_1$ and $T_2$ values found in the brain at 3T. Dictionary entries were simulated for $T_1$ values of [20:10:3000, 3200:200:5000] ms, $T_2$ values of [10:5:300, 350:50:500] ms, and 41 $B_1^+$ scaling factors between 0.5 and 1.5. Magnetization transfer and diffusion effects were neglected. The total time to calculate a dictionary for each flip angle pattern was approximately 1 hour and 45 minutes. 

\subsection{MRF Pulse Sequence and Acquisition}
A three-dimensional MRF FISP sequence was implemented on a 3T system (MAGNETOM Verio, Siemens Healthineers, Erlangen, Germany). A stack-of-stars k-space trajectory was used for data encoding. A slab-selective excitation was used for the FISP data collection, and a non-selective adiabatic hyperbolic secant pulse was used for inversion preparation. For each repetition of the MRF pattern, a single Cartesian k-space partition along the slab-select direction was acquired. The radial spokes rotated by the golden angle\cite{Winkelmann2007} ($111.246^\circ$) between adjacent FISP repetition times. The FISP TR was set to 8 ms, and the time between adiabatic inversion pulses was 10 s with an inversion time of 20 ms. The matrix size was fixed at 320 $\times$ 320 and 1 mm$^2$ voxels were acquired in-plane. The slab thickness was 168 mm and the slice thickness was 3.5 mm. Thus, 48 phase encoding partitions were acquired resulting in a scan time of 8 min for each MRF acquisition. A standard 12-channel head coil was used for signal reception. A $B_1^+$ mapping scan using the actual flip angle imaging (AFI) method\cite{Yarnykh2010} was also acquired at a resolution of 4 mm$^3$ with TR$_1$ of 20 ms, TR$_2$ of 200 ms, and a flip angle of 45$^\circ$. The $B_1^+$ maps were quantized to the resolution of the calculated dictionary. Matching was performed using the SVD compressed dictionary\cite{Doneva2010}.

\subsection{MRF Image Reconstruction}
The raw MRF data were first pre-whitened\cite{Pruessmann1999} to remove inter-RF coil noise correlations. Coil sensitivity maps were estimated using ESPIRiT\cite{Uecker2014} from a gradient echo 3D Cartesian calibration pre-scan integrated into the MRF sequence. A low-dimensional representation of the signal evolution was made using SVD of each calculated dictionary. The temporal signal evolution of the reconstructed parameter maps were then constrained to this subspace\cite{Tamir2017}. Note that constraining the reconstructed images to span a low-dimensional subspace is well established and distinct from the efforts in this work to constrain the acquisition parameters to span a low-dimensional temporal subspace. An alternating direction method of multipliers (ADMM) algorithm\cite{Boyd2010} with locally low rank (LLR) regularization\cite{Zhang2015} was used to reconstruct a set of $K_{Recon}=5$ subspace coefficient images. Twenty ADMM iterations, each with 10 conjugate gradient iterations were used. See Tamir et. al. for a description of the subspace constrained reconstruction used here\cite{Tamir2017}. The LLR regularization parameter of 0.02 was chosen empirically. Anisotropic delays between the readout gradient waveforms and the start of data collection were corrected using the autocalibrating SAGE method\cite{jiang2018simultaneous}. All reconstructions were performed with a combination of MATLAB and Berkeley Advanced Reconstruction Toolbox (BART, v0.6.00, \url{https://mrirecon.github.io/bart/}) commands on a laptop with a 6-core 2.6GHz processor and 32GB of RAM. The BART options used to reconstruct the subspace coefficient images were as follows: 'bart pics -R L:7:7:.02 -m -i20 -C10 -u0.01'.

\subsection{Phantom Experiments}

MRF and gold standard $T_1$ and $T_2$ relaxometry data were acquired in a calibration phantom (Eurospin Test Object 5). Reference $T_1$ values were obtained by a non-linear fit to the magnitude of multi-TR single spin echo images with TRs of 50, 100, 250, 500, 1000, 1500, 2000, 2500, 3000, 4000, and 5000 ms. An echo time of 12 ms was used for the $T_1$ mapping. Reference $T_2$ values were obtained by a non-linear fit to the magnitude of multi-TE single spin echo images with echo times of 12, 32, 52, 92, 142, and 192 ms. A TR of 8000 ms was used for the $T_2$ mapping. MRF data were acquired as described above for the flip angle pattern from Jiang et. al.\cite{Jiang2015} and the seven flip angle patterns optimized herein. Regression analyses compared the precision of the relaxometry values derived from MRF to the gold standard methods.



\section{Results}

\begin{figure}[h]
\centering
\includegraphics[width=1.0\textwidth]{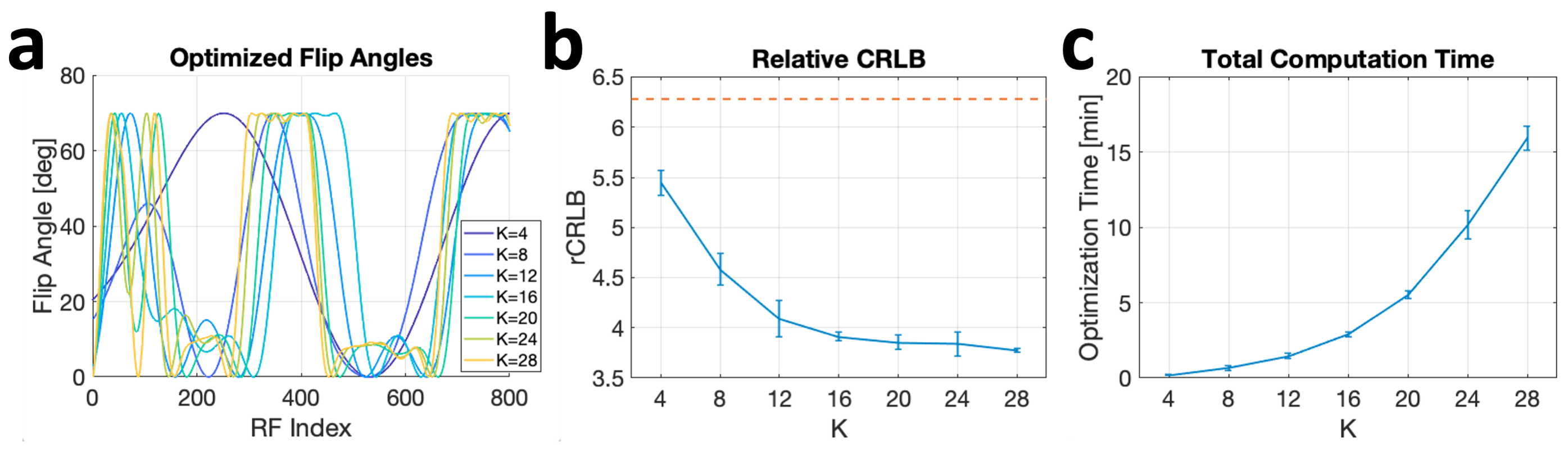}
\caption{(a) Optimized flip angle patterns for $K \in [4, 8, \dots, 28]$. (b) Mean relative Cramér-Rao Lower Bound (rCRLB) as a function of $K$. The dashed line here is the rCRLB of the flip angle pattern used in Jiang et. al. \cite{Jiang2015} (c) Mean total optimization time as a function of $K$. All error bars here represent the standard deviation.}
\label{fig:numopt}
\end{figure}

The numerical optimization results can be seen in Figure \ref{fig:numopt}. The rCRLB decreases as a function of $K$, but begins to plateau for $K \geq 16$. In all cases, the optimized pattern produces smaller rCRLB values than the original FISP MRF flip angle pattern, which is shown with the dashed line in Figure \ref{fig:numopt}b. The total calculation time increased exponentially with $K$ as seen in Figure \ref{fig:numopt}c. The optimal patterns shown in Figure \ref{fig:numopt}a tend to favor two properties: 1) the flip angle starts near zero, and 2) rises and falls between the maximum available range of flip angles, which is between $0^\circ$ and $70^\circ$ in this case. Despite being initialized with different random patterns, the solutions for $K \geq 16$ exhibit similar profiles, including plateaus at $70^\circ$ near RF indices 400 and 700.

\begin{figure}[h]
\centering
\includegraphics[width=0.75\textwidth]{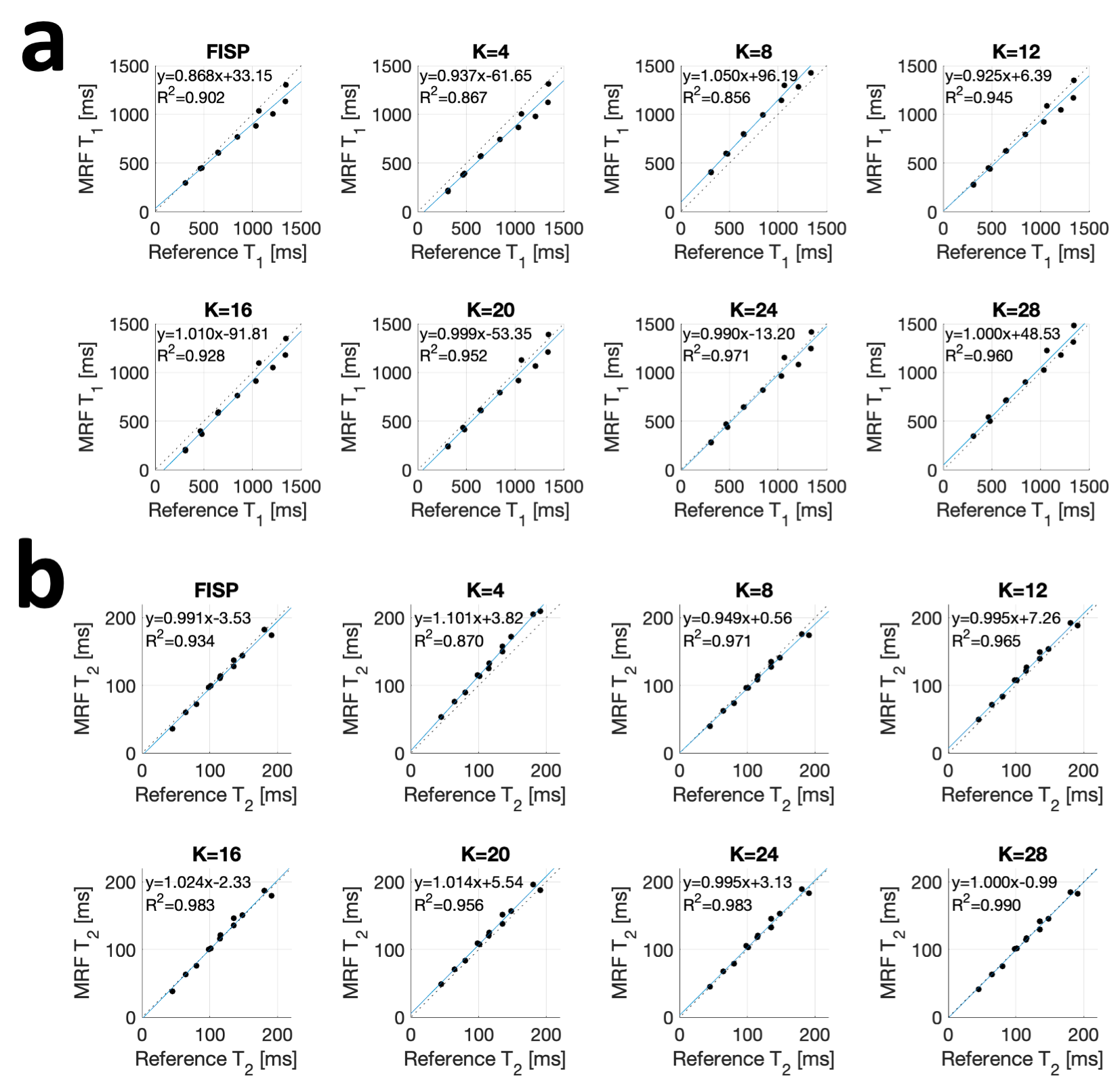}
\caption{(a) MRF-derived $T_1$ versus reference $T_1$ values for all tested flip angle patterns. (b) MRF-derived $T_2$ versus reference $T_2$ values for all tested flip angle patterns. In each plot, the dotted line represents the identity line and the solid line represents the lines of best fit used to calculate the $R^2$ values. }
\label{fig:regression}
\end{figure}

\begin{figure}[h]
\centering
\includegraphics[width=0.667\textwidth]{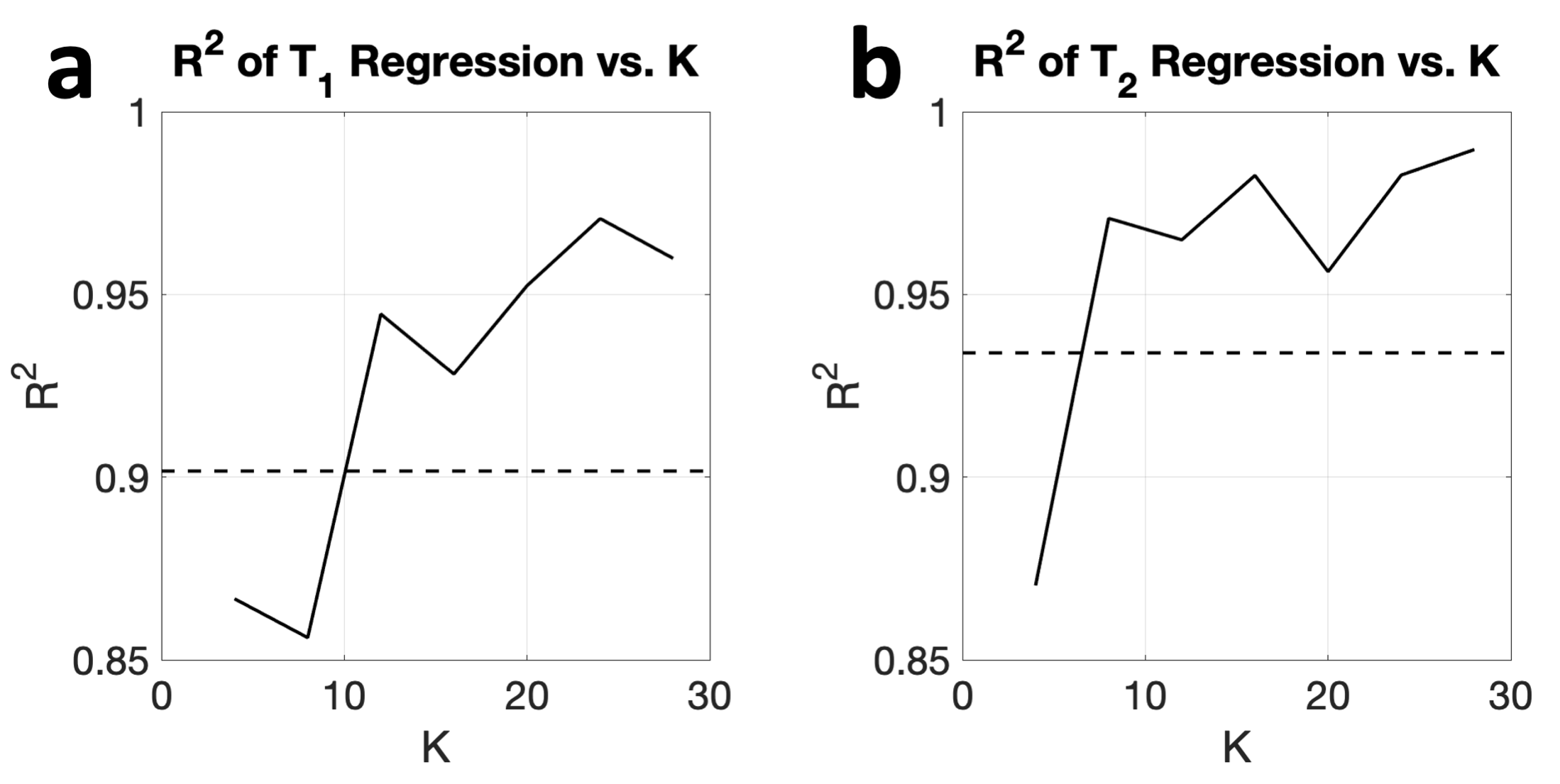}
\caption{Coefficient of determination ($R^2$) as a function of $K$ for (a) $T_1$ and (b) $T_2$. The dashed line shown in each plot represents the $R^2$ observed from the flip angle pattern of Jiang et. al.}
\label{fig:rmse}
\end{figure}

The results of the phantom experiments can be seen in Figures \ref{fig:regression} and \ref{fig:rmse}. Scatter plots showing the relationship between mean $T_1$ and $T_2$ values versus the single spin echo reference values are shown in Figure \ref{fig:regression}. The solid line in each plot represents the line of best fit, and the coefficient of determination ($R^2$) value for this fit is shown on each axis. For both $T_1$ and $T_2$, there is a very strong agreement between MRF-derived values and reference values as indicated by $R^2$ values near or well above 0.9. The $R^2$ values as a function of $K$ can be seen in Figure \ref{fig:rmse}. Since the coefficient of determination provides a value for the amount of variance in the MRF data explained by the reference data, we would expect $R^2$ to be higher for the optimized flip angle patterns. From the phantom experiments, $K \geq 12$ optimized coefficients were required to improve $R^2$ for $T_1$ fits, and $K \geq 8$ coefficients were required to improve $R^2$ for $T_2$ fits relative to the flip angle pattern of Jiang et. al. \cite{Jiang2015}.

\begin{figure}[h]
\centering
\includegraphics[width=1.0\textwidth]{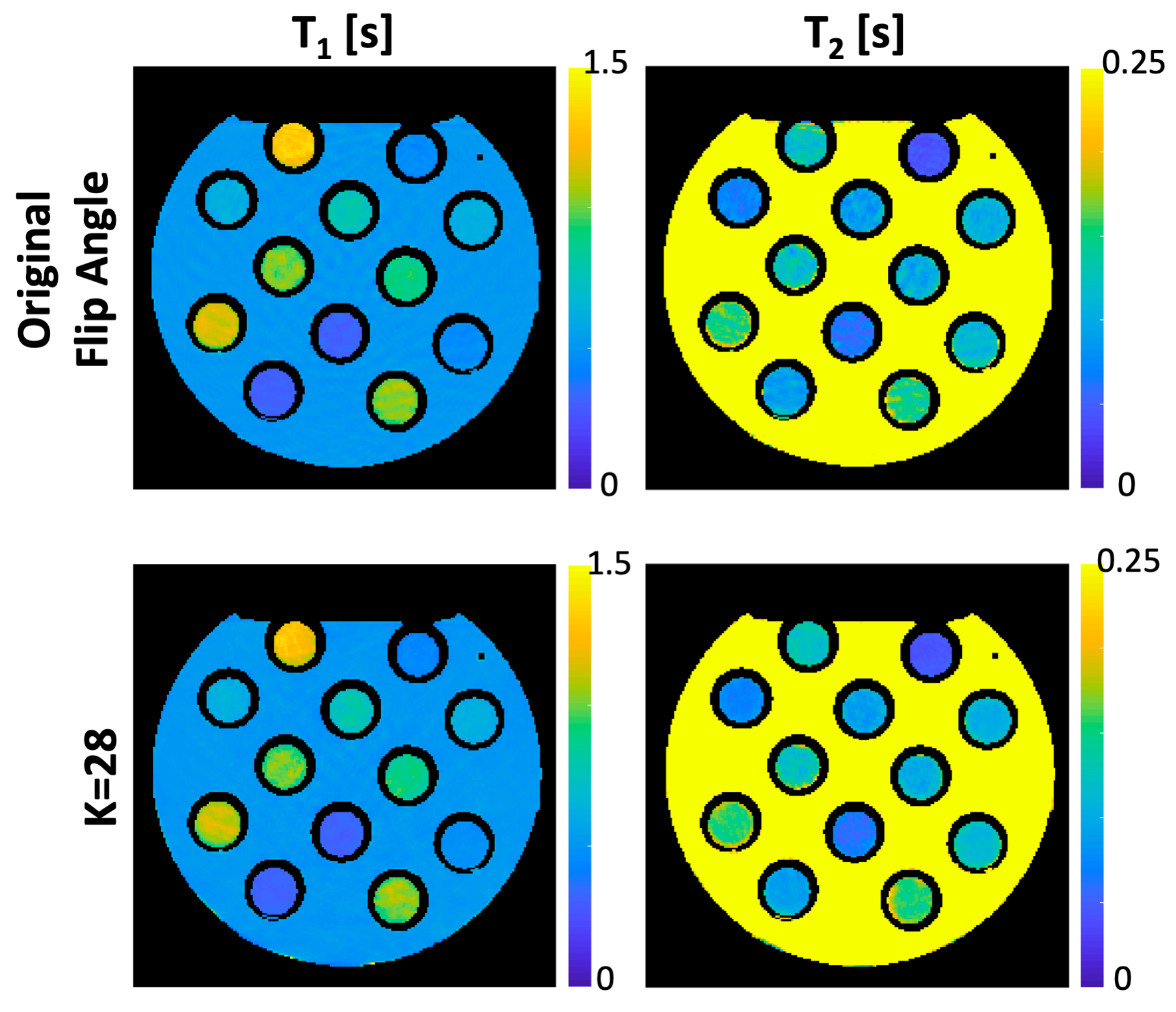}
\caption{Quantitative $T_1$ and $T_2$ maps from the original flip angle pattern for FISP MRF and the $K=28$ pattern optimized in this work to minimize the relative Cramér-Rao lower bound.}
\label{fig:phantom_images}
\end{figure}

Quantitative $T_1$ and $T_2$ maps from a central slice of the phantom experiments can be seen in Figure \ref{fig:phantom_images} for the flip angle pattern of Jiang et. al. and the $K=28$ pattern optimized in this work. Visually, a reduction in noise, intra-vial $T_1$ and $T_2$ variation, and streaking artifacts can be seen in the $K=28$ quantitative maps. Overall, the quality of the images appears better than the original flip angle pattern in addition to the improvements in addition to the improvements in the mean values within each of the phantom vials. 

\section{Discussion}

This work presents a method for optimal experiment design for magnetic resonance fingerprinting using a very small number of control parameters. The results demonstrate that more precise $T_1$ and $T_2$ values can be obtained by minimizing the Cramér-Rao lower bound of flip angle patterns represented with more than an order of magnitude fewer coefficients than there are RF pulses in the MRF acquisition. This optimization can be performed in less than five minutes. As such, it may be feasible to optimize flip angle patterns for many specific applications in a very short amount of time. We observed improvements in both $T_1$ and $T_2$ precision for $K \geq 12$ Gaussian basis function coefficients compared with the flip angle pattern from the original FISP MRF publication. 

While we focused on flip angle optimization in the present work, this framework could very easily be extended to simultaneously optimize repetition time, echo time, magnetization transfer off-resonance frequencies and RF power, etc. for various other MRF applications\cite{cohen2018rapid,jaubert2020water,Cloos2016}. In addition to MRF, such an optimization framework would also be applicable to $T_2$-shuffling\cite{Tamir2017} to maximize contrast between different tissue types from point-to-point throughout the long refocusing pulse train of 3D fast spin echo acquisitions. 

A number of Gaussian basis functions were used for a low dimensional representation of the flip angle train. This was chosen over polynomial or Fourier basis functions to maintain local control over the flip angle rather than having global functions effecting the entire flip angle train. An interesting avenue of future research would be a robust comparison of Gaussian, polynomial, Fourier, or data driven principal component analysis-derived basis sets in the context of MRF experiment optimization. 

While a full scale optimization (i.e., without low-dimensional constraints) may be beneficial from a theoretical standpoint despite enormous computation times, it would certainly never be feasible to optimize every flip angle individually based on a non-linear optimization of \textit{measured} errors in quantitative MRI parameters. In addition to speeding up \textit{in silico} MRF optimal experimental design, the framework presented here may also allow scanner-specific fine tuning to the flip angle pattern using a very small number of basis functions representing a slowly varying deviation from the theoretically optimal solution. Small system dependent imperfections such as $B_1^+$ transmit homogeneity, gradient performance, differences in $B_0$ homogeneity, and build up of eddy currents may cause the optimal MRF acquisition to vary from scanner-to-scanner. Thus, applying the proposed dimensionality reduction to the minimization of measured errors in $T_1$ and $T_2$ as a one-time calibration for each scanner could be performed. For $K=4$, approximately 150 evaluations of Equation \ref{eq:crb} were required for convergence. If a thin slice calibration phantom was used, deviations from the theoretically optimal flip angle pattern for a 3D FISP MRF sequence could be optimized to minimize measured errors in $T_1$ and $T_2$ in a reasonable amount of time (e.g., less than 2 hours). This is currently the subject of investigation.

Optimization of magnetic resonance fingerprinting experiments to improve the precision of quantitative parameter maps typically involves large scale optimizations that can take many hours to complete. Here, we presented and validated a method to reduce the dimensionality of such optimizations by representing the pattern of varying acquisition parmeters using a small number of basis functions. This allowed for a multi-start numerical optimization of flip angle patterns for FISP MRF that takes as little as 2 minutes to run while improving the precision of $T_1$ and $T_2$ maps compared to the conventionally used flip angle pattern.

\newpage 

\section{Appendix}
\subsection{Calculating Signal Evolution Derivatives for Inversion Recovery FISP MRF}

\subsubsection{Extended Phase Graph Review}
This section shows the calculations necessary to compute the partial derivatives shown in Equation \ref{eq:b_derivs} for the variable acquisition parameter schemes of an inversion recovery FISP MRF sequence using the extended phase graph (EPG) formalism\cite{Weigel2015}. The accuracy of performing a similar procedure to the following using Bloch equation simulations instead of EPGs depends on the number of isochromats used to simulate intravoxel dephasing. Many hundreds to thousands of isochromats may be required to obtain accurate results. The EPG formalism circumvents this issue by representing the MR signal in Fourier states, rather than as a numerical integration of magnetization at discrete sub-voxel locations. Furthermore, the implementation of the optimization framework in the EPG regime lays the foundation for simple incorporation of other effects such as diffusion or magnetization transfer\cite{Weigel2010,malik2018extended}. The configuration state matrix at radiofrequency (RF) pulse index $n$ is defined as follows.

\begin{equation}
    \mathbf{\Omega}_n=
    \begin{bmatrix}
        \widetilde{F}_{0,n} & \widetilde{F}_{1,n} & \widetilde{F}_{2,n} & \widetilde{F}_{3,n} & \dots \\
        \widetilde{F}^*_{0,n} & \widetilde{F}^*_{-1,n} & \widetilde{F}^*_{-2,n} & \widetilde{F}^*_{-3,n} & \dots \\
        \widetilde{Z}_{0,n} & \widetilde{Z}_{1,n} & \widetilde{Z}_{2,n} & \widetilde{Z}_{3,n} & \dots \\
    \end{bmatrix}
\end{equation}

The application of an RF pulse with a specified flip angle $\alpha_n$ and phase $\phi_n$ has the effect of "mixing" the configuration states. The transition operator, $T_n$, models the the effects of an instantaneous RF pulse and acts on the configuration matrix as below. The superscripts, - and +, denote the state immediately before and after the application of the RF pulse, respectively.

\begin{equation}
    \mathbf{\Omega}^+_n=\mathbf{T}\left(\alpha_n,\phi_n\right) \mathbf{\Omega}^-_n
\end{equation}

\begin{equation} 
    \mathbf{T}\left(\alpha_n,\phi_n \right) = 
    \begin{bmatrix}
        cos^2\frac{\alpha_n}{2} & e^{i2\phi_n}sin^2\frac{\alpha_n}{2} & -ie^{i\phi_n}sin{\alpha_n} \\
        e^{-i2\phi_n}sin^2\frac{\alpha_n}{2} & cos^2\frac{\alpha_n}{2} & ie^{-i\phi_n}sin{\alpha_n} \\
        \frac{-i}{2}e^{-i\phi_n}sin\alpha_n & \frac{i}{2}e^{i\phi_n}sin\alpha_n & cos\alpha_n
    \end{bmatrix}
\end{equation}

Transverse and longitudinal relaxation is also simply modeled using EPGs. Exponential decays due to $T_1$ and $T_2$ are modeled in the following matrix, $\mathbf{R}$.

\begin{equation}
    \mathbf{R}\left(t,T_1,T_2 \right) = 
    \begin{bmatrix}
        e^{-t/T_2} & 0 & 0 \\
        0 & e^{-t/T_2} & 0 \\
        0 & 0 & e^{-t/T_1}
    \end{bmatrix}
\end{equation}

Additionally, longitudinal magnetization recovery in the k=0 state is modeled with the matrix, $\boldsymbol{B}$.

\begin{equation}
    \mathbf{B}\left(t,T_1,M_0 \right) =
    \begin{bmatrix}
        0 & 0 & 0 & 0 & \dots \\
        0 & 0 & 0 & 0 & \dots \\
        M_0\left(1-e^{-t/T_1}\right) & 0 & 0 & 0 & \dots \\
    \end{bmatrix}
\end{equation}

One of the key benefits of EPGs is the simplicity in modeling the dephasing effect of spoiler or crusher gradients. The dephasing operation is carried out as a shifting, $\mathbf{S}$, of the values within the configuration state matrices. Note that the longitudinal states are not effected by the shifting operation.

\begin{equation}
    \mathbf{S}\left(\Delta k\right):\widetilde{F}_{k,n}\rightarrow\widetilde{F}_{k+\Delta k,n}, \widetilde{Z}_{k,n}\rightarrow\widetilde{Z}_{k,n}
\end{equation}

\subsubsection{EPG Representation of Inversion Recovery FISP MR Fingerprinting}
Using the EPG building blocks defined above, the effects of different pulse sequences can be modeled. Following a perfect adiabatic inversion pulse, the initial configuration state matrix immediately prior to the application of the first RF pulse in the FISP MRF pulse sequence is as follows, where $TI$ is the inversion time.

\begin{equation}\label{eq:initial_O}
    \mathbf{\Omega}_0=
    \begin{bmatrix}
        0 & 0 & 0 & 0 & \dots \\
        0 & 0 & 0 & 0 & \dots \\
        -M_0 e^{\frac{-TI}{T_1}} & 0 & 0 & 0 & \dots \\
    \end{bmatrix}
\end{equation}

The initial derivatives of $\mathbf{\Omega}_0$ with respect to each of the parameters of interest are:

\begin{equation}\label{eq:initial_dOdT1}
    \frac{\partial \mathbf{\Omega}_0}{\partial T_1}=
    \begin{bmatrix}
        0 & 0 & 0 & 0 & \dots \\
        0 & 0 & 0 & 0 & \dots \\
        \frac{M_0 \cdot TI}{T_1^2} & 0 & 0 & 0 & \dots
    \end{bmatrix}
\end{equation}

\begin{equation}\label{eq:initial_dOdT2}
    \frac{\partial \mathbf{\Omega}_0}{\partial T_2}=\mathbf{0}
\end{equation}

\begin{equation}\label{eq:initial_dOdM0}
    \frac{\partial \mathbf{\Omega}_0}{\partial M_0}=
    \begin{bmatrix}
        0 & 0 & 0 & 0 & \dots \\
        0 & 0 & 0 & 0 & \dots \\
        -e^{\frac{-TI}{T_1}} & 0 & 0 & 0 & \dots \\
    \end{bmatrix}
\end{equation}

The configuration state matrix at echo FISP TR index $n \in [1,\dots , N_{RF}]$ is given by the following recursive equation for FISP MRF.

\begin{equation}\label{eq:FISP}
    \mathbf{\Omega}_n := \mathbf{\Omega}^-_{n+1}=
    \mathbf{S}\cdot \left[ \mathbf{R}(TR_n,T_1,T_2)\mathbf{T}(\alpha_n,\phi_n)\mathbf{\Omega}_{n-1} + \mathbf{B}(TR_n,T_1,T_2) \right]
\end{equation}

Since the shifting operation does not affect longitudinal states, $\mathbf{S}\cdot \mathbf{B}(TR_n,T_1,M_0) = \mathbf{B}(TR_n,T_1,M_0)$. Therefore, Equation \ref{eq:FISP} can be simplified to the following.

\begin{equation}\label{eq:FISP_reduced}
    \mathbf{\Omega}_n =
    \mathbf{S}\cdot \mathbf{R}(TR_n,T_1,T_2)\mathbf{T}(\alpha_n,\phi_n)\mathbf{\Omega}_{n-1} + \mathbf{B}(TR_n,T_1,T_2)
\end{equation}

The signal at the echo time, $TE_n$, can be calculated as follows for FISP
\begin{equation}\label{eq:signal}
    \widetilde{F}_{0,n}=
    \mathbf{P}\left[\mathbf{R}(TE_n,T_1,T_2)\mathbf{T}(\alpha_n,\phi_n)\mathbf{\Omega}_{n-1} + \mathbf{B}(TE_n,T_1,M_0) \right]
\end{equation}
where $\mathbf{P}$ is an operator that extracts the rephased echo pathway, $\tilde{F}_0$, which is the complex-valued number at the first row and first column index of a state configuration matrix.
\begin{equation}
    \mathbf{P}\mathbf{\Omega}=(\mathbf{\Omega})_{1,1}
\end{equation}
Since $\mathbf{P}\mathbf{B}(TE_n,T_1,M_0)=0$, Equation \ref{eq:signal} can be reduced to the following.
\begin{equation}\label{eq:signal_reduced}
    \widetilde{F}_{0,n}=
    \mathbf{P}\mathbf{R}(TE_n,T_1,T_2)\mathbf{T}(\alpha_n,\phi_n)\mathbf{\Omega}_{n-1}
\end{equation}

\subsubsection{Partial Derivative With Respect to Longitudinal Relaxation Rate}

Taking the derivative of Equation \ref{eq:signal_reduced} with respect to $T_1$ via the product rule yields the following.
\begin{equation}\label{eq:dF0_dT1}
    \frac{\partial \widetilde{F}_{0,n}}{\partial T_1}=
    \mathbf{P}\frac{\partial \mathbf{R}(TE_n,T_1,T_2)}{\partial T_1}
    \mathbf{T}(\alpha_n,\phi_n) \mathbf{\Omega}_{n-1} + 
    \mathbf{P}\mathbf{R}(TE_n,T_1,T_2) \mathbf{T}(\alpha_n,\phi_n) \frac{\partial \mathbf{\Omega}_{n-1}}{\partial T_1}
\end{equation}
where
\begin{equation}
    \frac{\partial \boldsymbol{R}(TE_n,T_1,T_2)}{\partial T_1}=
    \begin{bmatrix}
        0 & 0 & 0 \\
        0 & 0 & 0 \\
        0 & 0 & \frac{TE_n}{T^2_1} e^{-TE_n/{T_1}}
    \end{bmatrix}
\end{equation}
Since $\mathbf{P}\frac{\partial \mathbf{R}(TE_n,T_1,T_2)}{\partial T_1}=0$, Equation \ref{eq:dF0_dT1} be simplified to:
\begin{equation}\label{eq:dF0_dT1_reduced}
    \frac{\partial \widetilde{F}_{0,n}}{\partial T_1}=
    \mathbf{P}\mathbf{R}(TE_n,T_1,T_2) \mathbf{T}(\alpha_n,\phi_n) \frac{\partial \mathbf{\Omega}_{n-1}}{\partial T_1}
\end{equation}

Taking the partial derivative of Equation \ref{eq:FISP_reduced} with respect to $T_1$ yields the following. 
\begin{equation}\label{eq:dFISP_dT1}
    \frac{\partial \mathbf{\Omega}_n}{\partial T_1}=
    \mathbf{S} \cdot \frac{\partial \mathbf{R}(TR_n,T_1,T_2)}{\partial T_1}
    \mathbf{T}(\alpha_n,\phi_n)\mathbf{\Omega}_{n-1} + 
    \mathbf{S} \cdot \mathbf{R}(TR_n,T_1,T_2) \mathbf{T}(\alpha_n,\phi_n) 
    \frac{\partial \mathbf{\Omega}_{n-1}}{\partial T_1} + 
    \frac{\partial \mathbf{B}(TR_n,T_1,M_0)}{\partial T_1}
\end{equation}
where
\begin{equation}
    \frac{\partial \mathbf{R}(TR_n,T_1,T_2)}{\partial T_1} =
    \begin{bmatrix}
        0 & 0 & 0 \\
        0 & 0 & 0 \\
        0 & 0 & \frac{TR_n}{T^2_1} e^{-TR_n/{T_1}}
    \end{bmatrix}
\end{equation}
and 
\begin{equation}
    \frac{\partial \mathbf{B}(TR_n,T_1,M_0)}{\partial T_1}=
    \begin{bmatrix}
        0 & 0 & 0 \\
        0 & 0 & 0 \\
        \frac{-M_0 \cdot TR_n}{T_1^2}e^{-TR_n/{T_1}}  & 0 & 0
    \end{bmatrix}
\end{equation}
Equations \ref{eq:dF0_dT1_reduced} and \ref{eq:dFISP_dT1} are then solved recursively given the initial conditions in Equations \ref{eq:initial_O} and \ref{eq:initial_dOdT1}.

\subsubsection{Partial Derivative With Respect to Transverse Relaxation Rate}
Taking the partial derivative of Equation \ref{eq:signal_reduced} with respect to $T_2$ gives the following. 
\begin{equation}\label{eq:dF0_dT2}
    \frac{\partial \widetilde{F}_{0,n}}{\partial T_2}=
    \mathbf{P}\frac{\partial \mathbf{R}(TE_n,T_1,T_2)}{\partial T_2}
    \mathbf{T}(\alpha_n,\phi_n)\mathbf{\Omega}_{n-1} + 
    \mathbf{P}\mathbf{R}(TE_n,T_1,T_2)\mathbf{T}(\alpha_n,\phi_n)
    \frac{\partial \mathbf{\Omega}_{n-1}}{\partial T_2}
\end{equation}
where
\begin{equation}
    \frac{\partial \mathbf{R}(TE_n,T_1,T_2)}{\partial T_2}=
    \begin{bmatrix}
        \frac{TE_n}{T_2^2}e^{-TE_n/T_2} & 0 & 0 \\
        0 & \frac{TE_n}{T_2^2}e^{-TE_n/T_2} & 0 \\
        0 & 0 & 0
    \end{bmatrix}
\end{equation}
Taking the partial derivative of Equation \ref{eq:FISP_reduced} with respect to $T_2$ yields
\begin{equation}\label{eq:dFISP_dT2}
    \frac{\partial \mathbf{\Omega}_n}{\partial T_2}=
    \mathbf{S}\cdot\frac{\partial \mathbf{R}(TR_n,T_1,T_2)}{\partial T_2}
    \mathbf{T}(\alpha_n,\phi_n)\mathbf{\Omega}_{n-1} +
    \mathbf{S} \cdot \mathbf{R}(TR_n,T_1,T_2) \mathbf{T}(\alpha_n,\phi_n)
    \frac{\partial \mathbf{\Omega}_{n-1}}{\partial T_2}
\end{equation}
where
\begin{equation}
    \frac{\partial \mathbf{R}(TR_n,T_1,T_2)}{\partial T_2} = 
    \begin{bmatrix}
        \frac{TR_n}{T_2^2}e^{-TR_n/T_2} & 0 & 0 \\
        0 & \frac{TR_n}{T_2^2}e^{-TR_n/T_2} & 0 \\
        0 & 0 & 0
    \end{bmatrix}
\end{equation}
Equations \ref{eq:dF0_dT2} and \ref{eq:dFISP_dT2} are solved recursively using the initial conditions in Equations \ref{eq:initial_O} and \ref{eq:initial_dOdT2}.

\subsubsection{Partial Derivative With Respect to Proton Density}

Taking the partial derivative of Equation \ref{eq:signal_reduced} with respect to $M_0$ yields the following.
\begin{equation}\label{eq:dF0_dM0}
    \frac{\partial \widetilde{F}_{0,n}}{\partial M_0}=
    \mathbf{P} \mathbf{R}(TE_n,T_1,T_2) \mathbf{T}(\alpha_n,\phi_n) 
    \frac{\partial \mathbf{\Omega}_{n-1}}{\partial M_0}
\end{equation}
Taking the derivative of Equation \ref{eq:FISP_reduced} with respect to $M_0$ produces
\begin{equation}\label{eq:dFISP_dM0}
    \frac{\partial \mathbf{\Omega}_n}{\partial M_0}=
    \mathbf{S} \cdot \mathbf{R}(TR_n,T_1,T_2) \mathbf{T}(\alpha_n,\phi_n)
    \frac{\partial \mathbf{\Omega}_{n-1}}{\partial M_0} +
    \frac{\partial \mathbf{B}(TR_n,T_1,M_0)}{\partial M_0}
\end{equation}
where
\begin{equation}
    \frac{\partial \mathbf{B}(TR_n,T_1,M_0)}{\partial M_0}=
    \begin{bmatrix}
        0 & 0 & 0 \\
        0 & 0 & 0 \\
        1-e^{-TR_n/T_1} & 0 & 0
    \end{bmatrix}
\end{equation}

Equations \ref{eq:dF0_dM0} and \ref{eq:dFISP_dM0} are solved recursively using the initial conditions shown in Equations \ref{eq:initial_O} and \ref{eq:initial_dOdM0}.

\subsubsection{Vectors for rCRB Optimization}
The vectors needed to calculate the Fisher information matrix as in Equation \ref{eq:b_derivs} are obtained by stacking the partial derivative values with respect to each parameter, $\theta_i$ into the corresponding vector, where $\theta=[T_1, T_2, M_0]^T$ and $i \in [1, 2, 3]$.
\begin{equation}
    \frac{\partial \widetilde{\mathbf{F}}_0}{\partial \theta_i}=
    \begin{bmatrix}
    \frac{\partial \widetilde{F}_{0,1}}{\partial \theta_i} & 
    \frac{\partial \widetilde{F}_{0,2}}{\partial \theta_i} &
    \dots &
    \frac{\partial \widetilde{F}_{0,N_{RF}}}{\partial \theta_i} 
    \end{bmatrix}^T
\end{equation}

\bibliography{references}

\end{document}